\def\eps{{\em Earth Planet. Space}}

\def\nat{{\em Nature}}
\def\natphy{{\em Nature Physics}}

\def\nu{{\em Nuclear Fusion}}
\def\prl{{\em Phys. Rev. Lett.}}

\def\pof{{\em Phys. Fluid}}

\documentclass[twocolumn]{aastex61}
\usepackage{natbib}
\usepackage{amsmath}

\newcommand\aastex{AAS\TeX}

\received{\today}
\submitjournal{ApJ Letter}

\shorttitle{\aastex\ Magnetic Reconnection Rate}
\shortauthors{H. Che}

\begin{document}

\title{ On the Rates of Steady, Quasi-steady and Impulsive Magnetic Reconnection }



\author{H. Che}
\affiliation{University of Maryland, College Park, MD, 20742, USA }
\affiliation{ NASA Goddard Space Flight Center, Greenbelt, MD, 20771, USA}

\begin{abstract}
Magnetic reconnection (MR) is considered as an important mechanism for particle energization in astrophysical plasma. Analyses of MR often assume the magnetostatic condition, i.e. $\partial_t = 0$, but various studies have concluded that MR cannot be steady. Using Maxwell and Poynting equations, we show: 1) Under the Sweet-Parker-Petschek framework, magnetostatic conditions produce contradictory results suggesting steady state cannot be achieved. In addition, fast MR must be compressible and magnetic flux is not conserved; 2) The quasi-steady MR defined as reconnection electric field being constant, i.e., $\partial_t \mathbf{E} = 0$, but $\partial_t \mathbf{B} \neq 0$ or equivalently $\partial_t\mathbf{j}\neq 0$, better describes the asymptotic behavior of non-turbulent Petschek-like MR. The conservation of mean Poynting flux implies that a fast MR does not require strong dissipation in the diffusion region. The upper limit of MR rate for quasi-steady MR is found to be $\sim 1/3\sqrt{3} \sim 0.2$.  3) For impulsive MR ($\partial_t\mathbf{B}_r \neq 0$ or $\partial_t\mathbf{j}_r \neq 0$ and $\partial_t\mathbf{E}_r \neq 0$), the MR rate is not bounded by the limit found for quasi-steady MR. The impulsive MR rate can be higher or lower than $1/3\sqrt{3}$ depending on factors such as the evolution stages of the MR and turbulence. Our analysis is independent of mass ratio and dissipation mechanism, thus the above conclusions can be applied to MR in pair plasma.
\end{abstract}

\keywords{magnetic reconnection --- acceleration of particles --- plasmas---turbulence}

\section{Introduction} 
\label{intro}
Magnetic reconnection (MR) is believed to be an important mechanism for particle energization in magnetospheric substorms \citep{baker96jgr,zelenyi10}, solar wind \citep{zank14apj}, and solar flares \citep{benz17lrsp}, and is drawing increasing interests in its possible roles in astrophysical phenomena such as the origin of the solar wind \citep{gloeckler03jgr, fisk03jgr}, $\gamma$-ray flares in the Crab Nebula \citep{buhler14,blandford17ssr} or pulsar nebulae in general, and $\gamma$-ray bursts \citep{grb12book,blandford17ssr}.   

From the beginning, the study of MR is dominated by two independent approaches \citep{sonnerup79conf, biskamp93book}: one considers the driven steady MR, which refers to open, externally forced reconnections \citep{sweet58book,parker57jgr,petschek64conf,sonnerup88,biskamp93book},  while the second approach concerns unsteady spontaneous MRs that arise from internal current instabilities whose dynamical evolutions only weakly depend on the external coupling \citep{dungey61prl,coppi66prl,galeev79ssr}. The steady MR approach has attracted wide interests since in many systems, the size of the reconnection region is much smaller than the spatial scale of the system. The coupling between the reconnection region and the external system occurs through the boundary condition imposed on the subsystem. In the collisional Sweet-Parker \citep{sweet58book,parker57jgr} and Petschek \citep{petschek64conf} models, the small region where the ideal MHD frozen-in condition $\mathbf{E}+\mathbf{U}\times\mathbf{B}/c=0$ breaks is called the \textit{diffusion region} (DR). 
Later the steady MR model is expanded to include non-collisional terms in the generalized Ohm's law that break the frozen-in condition \citep{vasyliunas75rg,sonnerup88,gurnett05book,che11nat}. These non-ideal terms include the non-gyrotropic pressure gradient, the convective momentum transport, the Hall effect, and the anomalous dissipation due to kinetic-scale turbulence. 

The fundamental issue in MR is how to achieve the fast magnetic energy conversion seen in observations. The generalized Sweet-Parker and Petschek models offer the theoretical framework to address the problem. The normalized reconnection rate, defined as $R \equiv U_I/c_A$, where $U_I$ is the speed of inflow plasma from the external system into the diffusion region, and $c_A$ is the Alfv\'en speed. Constraints of reconnection rate come largely from numerical simulations, particularly particle-in-cell (PIC) simulations. Some simulations seem to suggest that fast collisionless MR is controlled by Hall effect and has a universal rate 0.1 \citep{shay99grl}. Other simulations show that anomalous effects can accelerate MR processes to be faster than the Hall MR rate \citep{bhat99jgr,che11nat,che17pop,munoz17} but not always \citep{daughton11natphy,le18pop}. In relativistic pair plasma simulations in which hall effect is zero due to the equal mass of particles, $U_I/c$ can reach as high as 0.6 (the relativistic $c_A$ is smaller than $c$)\citep{blandford17ssr,papini18}.   

Direct observations of the plasma inflow show that MRs in solar flares are unsteady and the rates vary from 0.01-0.5 \citep{su13nat}, while indirect measurements of the inflow using the motion of magnetic flux tubes at the foot-points of magnetic loops, assuming the MRs being steady and the magnetic fluxes conserved, found the MR rates $<$ 0.1 \citep{qiu02apj,qiu04apj}. A large number of unsteady reconnections called fast flux transfer events (FTEs), have been discovered in the magnetopause since 1970s \citep{russell78ssr}. \textit{In situ} {\it Magnetospheric Multiscale Science} (MMS) observations of  the magnetopause MR events show that the reconnection rate can be $>$ 0.1 facilitated by anomalous effects \citep{torbert17jgr}. The rates for impulsive MR in laboratory plasma lie in a large range varying from 0.01 to $>$0.5 \citep{fox11prl,dorfman12phd}.

How to reconcile these seemingly controversial results is a profound challenge to our understanding of MR.  Studies on MR rate often assumes the magnetostatic condition, i.e. $\partial_t = 0$. However, rigorous calculations have shown that steady solutions in the DR, and the boundary conditions can not be self-consistently obtained in the Sweet-Parker model \citep{biskamp93book}, and the Petschek-like MR is intrinsically not steady \citep{syro71jepp,zelenyi10}. \citet{kulsrud01eps} showed that Petschek MR is equivalent to Sweet-Parker MR if steady condition is imposed, implying MR can not be steady. This has been demonstrated in resistive MHD numerical simulations, which show that steady Sweet-Parker MR is unrealistic \citep{birn01jgra}, and Petschek-like MR can only be achieved when the resistivity and electric field are centralized near the null-point \citep{sato79pof,birn01jgra}. It is obvious that a non-uniform $E_z$ requires $\partial_t B \neq 0$.

In this letter using Maxwell and Poynting equations, we demonstrate that under the Sweet-Parker-Petschek (SPP) framework, steady MR ansatz causes contradictory results, indicating that MR is not intrinsically steady. Quasi-steady MR, defined as reconnection electric field $\partial_t \mathbf{E_r} = 0$ but reconnection magnetic field $\partial_t \mathbf{B_r} \neq 0$ (or equivalently the associated current density $\partial_t\mathbf{j_r}\neq 0$), better describes the ``asymptotic" behavior of non-turbulent MR. The upper limit of reconnection rate for quasi-steady MR is found to be $1/3\sqrt{3}\approx 0.2$. In turbulent/impulsive reconnection where  $\partial_t\mathbf{B_r} \neq 0$ and $\partial_t\mathbf{E_r} \neq 0$, the reconnection rate can be higher than this limit as the magnetic flux piles up in DR. Our conclusions are independent of the mass ratio and dissipation processes in DR, and are applicable to pair plasma. 
\section{Reconnection rate} 
The Poynting equation 
\begin{equation}
\partial_t W +\mathbf{j}\cdot\mathbf{E}+\nabla\cdot \mathbf{N}=0,
\label{dpoyn}
\end{equation}
where $W=(B^2+E^2)/8\pi$ is the field energy,  $\mathbf{N}=c\mathbf{E}\times\mathbf{B}/4\pi$ is the Poynting vector and $\mathbf{j}\cdot\mathbf{E}$ the plasma heating, 
describes two essential processes in MR: the electromagnetic energy conversion in the DR and the transport of Poynting flux into and out of the DR. In this section we show that the Poynting equation provides a short-cut to constrain the reconnection rate without the need to consider the dissipation mechanism in the DR.  

\begin{figure}
\includegraphics[scale=0.3,trim=100 150 100 60,clip]{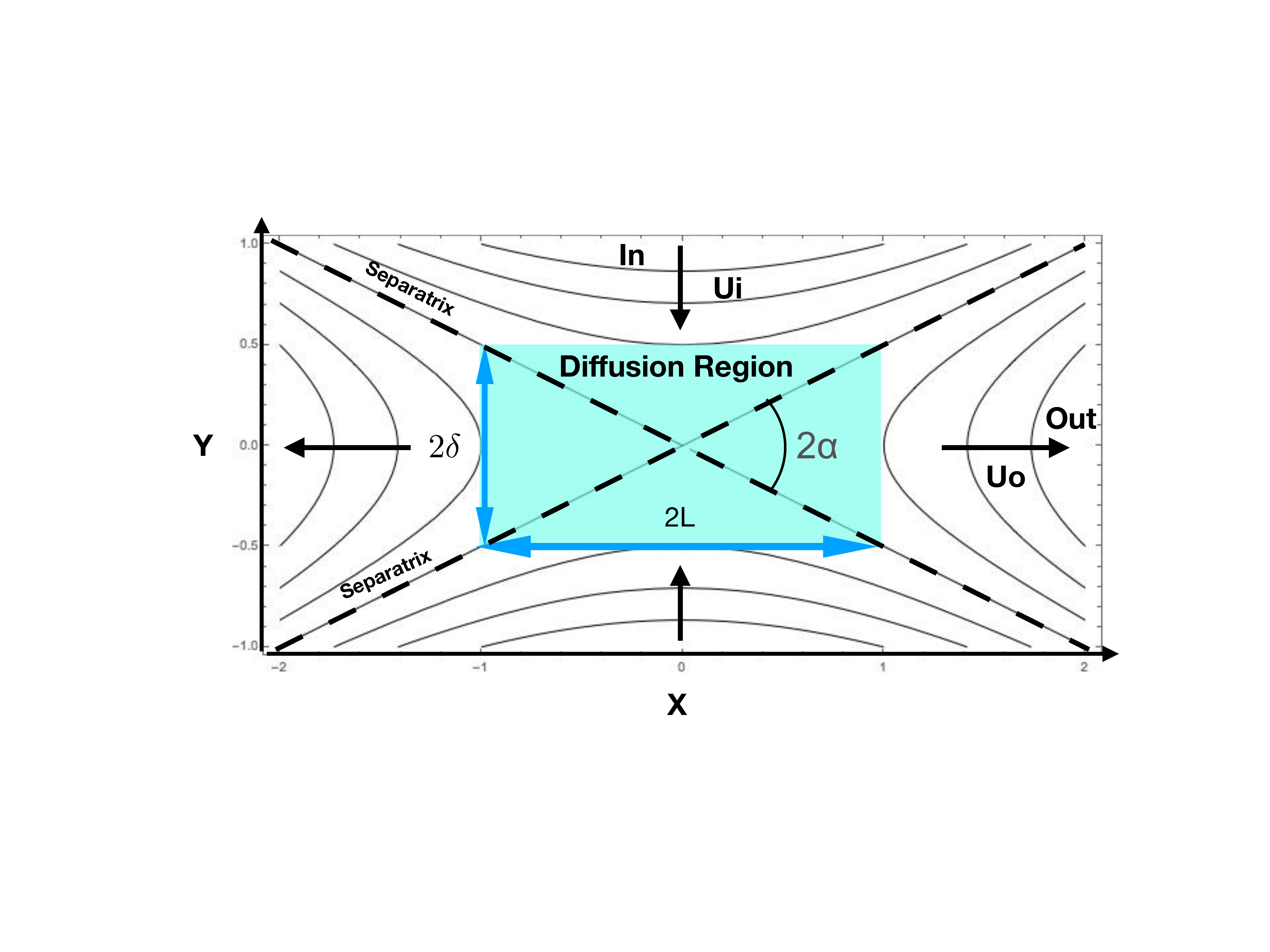}
\caption{An illustration of the magnetic field structure in the reconnection plane of the SPP MR. The solid lines represent the magnetic field lines. The separatrices cross the DR (dashed lines). The guide field $\mathbf{B}_g$ is in z direction and is not shown.}
\label{illu}
\end{figure}

Although MR in general is 3D, only the anti-parallel magnetic field components $\mathbf{B_r}$ are involved in the field annihilation, the guide-field $\mathbf{B_g}$ on the other hand may affect the physical processes inside the DR \citep{swisdak05jgr,sauppe18pop}. MR with a guide-field is also known as component reconnection  \citep{swisdak05jgr}. While the original SPP framework describes MR in 2D, it can be considered as a model for the reconnection of the anti-parallel component in a 3D MR. Following the common practice in space and astrophysical plasma \citep{vanb85apj,biskamp93book,yamada10romp,boozer18jpp}, let $\mathbf{B_r}= B_x\hat{x}+B_y\hat{y}$ be in the $xy$ plane (Fig.~\ref{illu}),  and $\mathbf{B_g}=B_g\hat{z}$ in the direction perpendicular to the reconnection plane and is a constant. SPP reconnection is characterized by an X-Type neutral point magnetic field geometry determined by $\nabla\cdot\mathbf{B}=0$ \citep{parnell96pop}. The DR can be approximated as a box with dimensions of $2\delta \times 2L$. Inside the DR the magnetic field lines break and reconnect as determined by the generalized Ohm's law \citep{vasyliunas75rg}. Outside the DR, the MHD ideal frozen-in condition $\mathbf{E}+\mathbf{U}\times \mathbf{B}/c=0$ holds and determines the transport of the magnetic flux into and out of the DR together with the plasma flow. The separatrices demarcate the inflow and outflow regions. We use subscripts/superscripts ``I" and ``O" to denote quantities in the inflow and outflow regions, respectively. The reconnecting magnetic field in the upper inflow region $\mathbf{B_r}$ is in the x-direction and $\mathbf{B_I}=B_I\hat{x}$, thus the magnetic field in the right outflow region points to the y-direction with $\mathbf{B_O}=B_O\hat{y}$.  From the frozen-in condition, we have in the upper inflow region $\mathbf{E}_z^I=-U_I B_I/c\hat{z}$ and in the right outflow region $\mathbf{E}_z^O=-U_OB_O/c\hat{z}$. 

\subsection{Can Magnetic Reconnection Be Steady ?}
\label{steady}
  
It is commonly assumed that after a fast onset phase MR can eventually reach a steady state,  i.e., $\partial_t = 0$, when the reconnection electric field peaks and the magnetic flux brought into the DR by the frozen-in plasma flow balances the merging of the magnetic field inside. This assumption implicitly excludes turbulent MR. In the following we show that the steady state ansatz can produce conflicting results, indicating magnetic reconnection cannot be steady.

$\partial_t \mathbf{B}_r= 0$ reduces the Faraday's law to $\nabla \times \mathbf{E} = 0$, or 
\begin{equation}
\label{far}
\partial_x E_z=0, \partial_y E_z =0.\\
\end{equation}
The steady Ampere's law becomes $\nabla\times\mathbf{B}=\frac{4\pi}{c}\mathbf{j}$. 
Obviously, the current is also steady, i.e.,
\begin{equation}
\partial_t \mathbf{j}_z= 0.
\end{equation}

Eq. (\ref{far}) implies that $E_z$ is a constant inside and outside the DR. Using the frozen-in condition, $E_z^I=E_z^O$ gives
\begin{equation}
U_IB_I=U_OB_O,
\label{conflux}
\end{equation}
i.e., the inflow and outflow magnetic fluxes are balanced. This implies that the magnetic flux is conserved \citep{newcomb58ap}, i.e.,  
\begin{equation}
\nabla\times (\mathbf{E}+\mathbf{U}\times\mathbf{B})=0.
\label{genflux}
\end{equation}
in the DR as the inflow magnetic fluxes move into the null region and out to the outflow region after the field-line reconnection.

Near the X-type neutral point with opening angle $\alpha$, it is easy to show 
\begin{equation}
\frac{B_O}{B_I}=\frac{\delta}{L}.
\label{rb}
\end{equation}
Combining with Eq.~(\ref{conflux}) we have
\begin{equation}
\frac{U_I}{U_O}=\frac{\delta}{L}.
\label{steadyu}
\end{equation}
Steady fluid equations imply that the magnetic pressure can accelerate outflow speed to $c_A$, and the reconnection rate $U_I/c_A\leq 1$. Eq.~(\ref{steadyu}) is the well-known scaling-law of the Sweet-Parker MR under the incompressible condition $\nabla\cdot\mathbf{U}=0$ \citep{sweet58book,parker57jgr}. In other words, steady state implies incompressibility.


Now we investigate the magnetic flux conservation, i.e., Eq.~(\ref{genflux}) inside the DR. Let's consider a small region inside the DR adjacent to the inflow boundary. The x-component of $\mathbf{U}$ and y-component of $\mathbf{B}$ are negligible, thus $\mathbf{U} = U_y \hat{y}$, $\mathbf{B} = B_x \hat{x}$.  $\nabla \cdot \mathbf{U}=0$ and $\nabla \cdot \mathbf{B} = 0$ reduce to 
\begin{gather}
\frac{\partial U_y}{\partial y} = 0,
\frac{\partial B_x}{\partial x} = 0.
\end{gather}
Expanding Eq.~(\ref{genflux}), and taking into account $\nabla \cdot \mathbf{B} = 0$, $\nabla \cdot \mathbf{U}=0$, and $\nabla \times \mathbf{E} =0$, we obtain:
\begin{equation}
B_x \frac{\partial U_y}{\partial x} \hat{y}- U_y \frac{\partial B_x}{\partial y} \hat{x} = 0.
\label{conflux2}
\end{equation}
In the DR, for $y \neq 0$, $B_x \neq 0$ and $U_y \neq 0$.  Therefore, 
\begin{gather}
\frac{\partial U_y}{\partial x} = 0, 
\frac{\partial B_x}{\partial y} = 0.
\end{gather}
Thus $U_y$ and $B_x$ must be non-zero constants. At the upper and lower boundaries of the DR, $B_x=B_I$, $U_y=U_I$, and $E_z^I$ is also a constant as shown earlier. This implies that the only solution to Eq.~(\ref{conflux2}) is the frozen-in condition $E_z^I+U_I\times B_I/c=0$, suggesting the small region should not be in the DR. Thus we need to redefine a smaller DR that does not include the small region we have carved out. Repeat this process and eventually the DR become infinitesimally small and we rule out the existence of the DR. This contradiction is clearly a consequence of the steady-state assumption.  Similar inconsistencies are found in solutions of MHD moment equations of steady MR \citep{syro71jepp, biskamp93book,zelenyi10}. That steady conditions prohibit MR is consistent with the resistive MHD MR simulations with a uniform resistivity whose rates are found to be consistently $\ll 0.1$ and negligible \citep{birn01jgra}.

  
A corollary of our results is that SPP MR is compressible and the magnetic flux is not conserved. 

\subsection{Reconnection rate of Quasi-steady Magnetic Reconnection}
\label{petschek}
 
We now relax the requirement of $\partial_t \mathbf{B} = 0$, and investigate the rate of quasi-steady non-turbulent Petschek-like MR that satisfies $\partial_t \mathbf{B} \neq 0$ or equivalently $\partial_t \mathbf{j} \neq 0$, but $\partial_t E_z = 0$.

From the Faraday's Law $\nabla\times\mathbf{E}=-\frac{1}{c} \partial_t \mathbf{B}$,  the mean $\overline{E}_z$ inside the DR is 
\begin{equation}
\overline{\mathbf{E}}_z=-\frac{\delta}{c}\frac{\triangle B}{\triangle t}\hat{z}, 
\label{er}
\end{equation}  
where $\overline{\mathbf{E}}_z$ is estimated at $y=\pm \delta/2$.
From $\nabla \times \mathbf{B}=\frac{4\pi}{c} \mathbf{j}$,  the mean current density is 
\begin{equation}
\overline{\mathbf{j}}_z=-\frac{c}{4\pi}\frac{B_I}{\delta}\hat{z},
\end{equation}
where we neglect the contributions from $E_x$, $j_x$, $E_y$ and $j_y$ associated with the spatial and temporal variations of $B_g$. Since $B_g$ does not participate in the field line merging and $\partial_t B_g\simeq 0$,  we then have
\begin{eqnarray}
\overline{\mathbf{j}}_z\overline{\mathbf{E}}_z =-\frac{c}{4\pi}\frac{B_I}{\delta} \frac{\delta}{c}\frac{\triangle B}{\triangle t}
=\frac{B_I\triangle B}{4\pi \triangle t}. 
\end{eqnarray}
The mean decrease of the electromagnetic energy $\partial \overline{W}/\partial t$ is approximately 
\begin{equation}
\frac{\triangle \overline{W}} {\triangle t}=-\frac{B_I\triangle B}{4\pi \triangle t}.
\end{equation}
Using $j_z  E_z\approx\overline{j}_z \overline{E}_z$, the Poynting equation is approximately $\nabla\cdot \overline{\mathbf{N}}=-\partial_t \overline{W}-\overline{j}_z \overline{E}_z$, 
and therefore inside the DR we have 
\begin{equation}
\nabla\cdot \overline{\mathbf{N}}=0.
\label{new_poyn}
\end{equation}
Eq.(\ref{new_poyn}) shows that quasi-steady reconnection is a rather delicate state in which the Poynting flux is conserved inside the DR. To achieve such a state the thermal dissipation in the DR must be small so that the annihilated magnetic field is regenerated through the increase of the electric current. Large thermal dissipation in the DR such as collisional resistivity or anomalous resistivity, on the other hand, impedes the increase of the current and tips the balance of the Poynting flux in the DR. 

The conservation of Poynting flux $\int_s \mathbf{N} \cdot d\mathbf{S} =  0$  yields:
\begin{equation}
\frac{\vert N_O\vert}{L} =\frac{\vert N_I\vert }{\delta}.
\label{Nflux}
\end{equation}
By definition 
\begin{equation}
 \mathbf{N}_O=-\frac{c}{4\pi}\mathbf{E}_z^O \mathbf{B}_O, 
 \mathbf{N}_I=\frac{c}{4\pi}\mathbf{E}_z^I \mathbf{B}_I. 
 \end{equation}
Then we get 
\begin{equation}
\frac{U_I}{U_O}=(\frac{\delta}{L})^3,
\end{equation}
where we used the frozen-in condition $\mathbf{E}_z^O=-B_OU_O/c\hat{z}$, $\mathbf{E}_z^I=-B_IU_I/c\hat{z}$, and $B_O/B_I=\delta/L$.

We now show that the mean reconnection rate and the aspect ratio of the DR $\delta/L$ can be estimated in quasi-steady MR. 

$\vert E_z\vert$ peaks in the midplane and decreases towards the boundary of the inflow region, thus $\vert E_z^O\vert$ along the midplane is close to the maximum of $\vert E_z\vert$ and $\vert E_z^I\vert$ is close to the minimum of $E_z$. We approximate $\overline{E}_z \approx (E_z^O + E_z^I)/2$, plug it into Eq.~(\ref{er}) and we have
\begin{equation}
\frac{B_I U_I+B_O U_O}{2c} \sim \frac{\delta}{c}\frac{\triangle B}{ \triangle t}. 
\end{equation} 
During $\Delta t$ the total change of magnetic field in the DR due to the annihilation of the anti-parallel $B_I$ is
$\triangle B=2B_I$,
and hence 
\begin{equation}
\frac{4\delta}{\triangle t}\sim U_I +\frac{B_O}{B_I}U_O =U_I+\frac{\delta}{L} U_O.
\end{equation}
Using the relation $\delta/\triangle t =U_I$ and $U_O/U_I=(L/\delta)^3$, we obtain
\begin{equation}
\frac{\delta}{L}\sim\frac{1}{\sqrt{3}},
\end{equation}
and the rate of Petschek-like MR  is 
\begin{equation}
\frac{U_I}{U_O}\sim\frac{1}{3\sqrt{3}}\approx 0.2.
\end{equation}
It should be noted that since the shrink of current sheet due to $\partial_t\mathbf{j}_r\neq 0$ costs part of the released magnetic energy, the ram pressure $mn_OU_O^2/2$ no longer balances the magnetic pressure $B^2/8\pi$ as in steady MR, and hence $U_O \leq c_A$. Thus the reconnection rate satisfies
\begin{equation}
R=\frac{U_I}{c_A}\leq \frac{1}{3\sqrt{3}}\approx 0.2.
\end{equation}

Finally we look at how variable the magnetic field is in quasi-steady MR. Using the frozen-in condition in the inflow region to replace $\overline{E}_z$ in Eq.~(\ref{er}), we obtain 
\begin{gather}
\frac{\triangle B/B_I}{\Omega_i \Delta t} \sim \frac{U_I/c_A}{\delta/d_i} \leq  \frac{0.2}{\delta/d_i},\\
\frac{\overline{E}_z/E_0}{\delta/d_i}\leq 0.2,
\label{prate}
\end{gather}
where $E_0=B_I c_A/c$, $d_i$ is the ion inertial length and $\Omega_i$ is the ion gyro-frequency. For a current sheet with width $\sim d_i$, the magnetic field varies by $\leq 20\%$ over $t \sim \Omega_i^{-1}$. The corresponding spatial gradient of $E_r$ is also $\leq 20\%$. 

%
%
%
\subsection{Unsteady Magnetic Reconnection}
 If $\partial_t\mathbf{B_r}\neq 0$ and  $\partial_t\mathbf{E_r}\neq 0$, MR becomes unsteady or impulsive, and the reconnection rate is not bounded by the limit we found for quasi-steady MR. Simulations of unsteady turbulent MR show that the rate can indeed exceed $1/3\sqrt{3} \sim 0.2$ \citep{che17pop,blandford17ssr}, but not all turbulent reconnections have high rates.  Under what circumstance could the reconnection rate exceed 0.2?  Let's consider a turbulent MR in which the mean field reaches a ``quasi-steady state", but some instabilities in the current sheet generate high-frequency waves, so that the MR is unsteady. In this case, we can split $\mathbf{E}$, $\mathbf{B}$, and $\mathbf{j}$ into the slow and fast changing parts, so that $\mathbf{E}=\langle\mathbf{E}\rangle +\delta \mathbf{E}$,  and $\langle \delta \mathbf{E}\rangle=0$, etc., where $\langle...\rangle$ represents the ensemble average. The turbulent part of the Poynting equation becomes
\begin{equation}
\partial_t \langle\delta W\rangle+ \langle\delta \mathbf{j}\cdot \delta \mathbf{E}\rangle+\nabla\cdot \langle\delta \mathbf{N}\rangle=0,
\label{turbp}
\end{equation}
 where $\langle \delta W\rangle=(\langle\delta\mathbf{B}^2\rangle+\langle\delta \mathbf{E}^2\rangle)/8\pi$, and $\langle\delta\mathbf{N}\rangle=c\langle\delta \mathbf{E}\times\delta\mathbf{B}\rangle/4\pi$.  In the DR, Anomalous turbulence effects generated by internal current instabilities, whose growth timescale is much shorter than the MR evolution timescale $\sim L/c_A$, enhance the magnetic field and non-thermal plasma heating by wave-particle interactions so that $\nabla\cdot \langle\delta\mathbf{N} \rangle<0$. Examples include anomalous resistivity produced by electrostatic instabilities \citep{yamada10romp,che17pop}, anomalous viscosity produced by electromagnetic instabilities such as electron velocity shear instability \citep{che11nat}. Since the mean-field is in a ``quasi-steady state" so that $\nabla\cdot \langle \mathbf{N} \rangle = 0$, we have $\nabla\cdot\mathbf{N} <0$ in the DR, consequently $U_I/U_O>(\delta/L)^3> 0.2$.

Clearly, anomalous turbulence effects do not necessarily result in high reconnection rate if the mean field MR does not reach a quasi-steady state or the turbulent enhancement is not strong enough. This is why some turbulent PIC MR simulations show anomalous effects significantly accelerates reconnection while others do not.

%
%

\section{Conclusions and discussions}
In this letter we revisited the rate of MR under the SPP framework. This model is particularly useful in open astrophysical environment. We show that the couplings between the reconnection electric field and the current, and the electromagnetic energy flux transfer determine the rate of MR. The main conclusions are:  1) The magnetostatic ansatz, i.e., $\partial/\partial t = 0$, leads to contradictory results for SPP MR. This suggests that the steady MR is not possible; 2) Steady state implies magnetic-flux conservation and incompressibility in the MR. A corollary of the first conclusion is that SPP MR is compressible and the magnetic flux is not conserved;  3) Non-turbulent Petschek-like MR can be quasi-steady, i.e., the reconnection electric field satisfies $\partial_t \mathbf{E}_r\sim 0$ but the reconnecting magnetic field $\partial_t \mathbf{B}_r\neq 0$ or the associated current density $\partial_t \mathbf{j}_r\neq 0$. The time variation of magnetic field is limited by $\Delta B/B_I/\Omega_i t \leq 0.2/\delta/d_i$. The characteristic of quasi-steady MR is the Poynting flux being nearly conserved in the DR, implying that the dissipation in the DR being small. The MR rate for quasi-steady MR is $U_I/c_A\leq (\delta/L)^3 \leq 1/3\sqrt{3} \sim 0.2$; 4) For impulsive MR driven by internal current instabilities in which $\partial_t \mathbf{B}_r\neq 0$ and $\partial_t \mathbf{E}_r\neq 0$, the rate can be higher or lower than  0.2. 
These results are applicable to both 2D and 3D MR. Guide field may affect the detailed processes in the DR which may affect the reconnection rate \citep{sauppe18pop}. However, the conclusions regarding steady, quasi-steady and unsteady MR should not change qualitatively.  Note that the equations in this analysis are intrinsically relativistic, and our analysis is independent of the mass ratio and the dissipation processes in the DR, thus the above conclusions are applicable to MR in relativistic pair plasma.

The near conservation of Poynting flux $\nabla\cdot\mathbf{N}\sim 0$ in quasi-steady MR means the dissipation inside the DR must be small, and the annihilated magnetic field is recovered by the reconnection electric field through the inertia $\partial_t \mathbf{j}_r \neq 0$.  Therefore, the width of the current sheet can not be constant. For example, collisionless MR can be fully supported by inertia without dissipation \citep{boozer18jpp}. The current sheet may shrink until it becomes unstable to instabilities driven by the intense magnetic/velocity shears, and subsequently the instabilities may broaden the current sheet. Various non-ideal effects, such as non-gyrotropic pressure and convective momentum transport \citep{ vasyliunas75rg,kuznetsova01jgr} may slow the narrowing of the current sheet on electron inertial scale, but cannot fully stabilize the current sheet since the reconnection electric field centralizes in the electron DR and globally is non-uniform. 
 
In resistive MR, $\nabla\cdot\mathbf{N}\sim 0$ implies a high Lundquist number $S \propto 1/\eta$, and when $S$ is larger than the corresponding critical value, tearing instability is triggered and the reconnection becomes impulsive \citep{loureiro16ppcf}. 

Impulsive MR behave like quasi-steady non-turbulent MR when the turbulence fully decays i.e. $\langle\delta N \rangle\sim 0$, or evolves into the fully developed state with the correlation scale comparable or larger than the size of the DR. In the latter case the turbulence effect is close to uniform spatially and thus $\nabla\cdot\langle\delta \mathbf{N} \rangle\sim 0$.

Solar flares are unsteady and commonly impulsive \citep{fletcher11ssr}. Assuming conservation of magnetic flux for such systems when measuring reconnection rate can underestimate the merging rate of magnetic field. This may explain the apparent discrepancy between \citet{su13nat} and \citet{qiu02apj, qiu04apj}.


\acknowledgments
HC would like to thank Roald Sagdeev for the constructive discussions. HC also likes to thank the helpful discussions with Russell Kulsrud on the anomalous resistivity, Lev Zeleny on the stability of MR; David Seibeck on the unsteady MR in magnetopause, Jiong Qiu and Brian Dennis on the observations of reconnection rate in solar flares, Joachim Birn and Michael Hesse on the compressible MHD simulations of MR. HC also thanks the anonymous referees for the insightful comments that help to improve the clarity of this manuscript. HC is partly supported by NASA grant No. NNX17AI19G and MMS project. 

\end{document}